\documentclass[twocolumn,aps, prl,showpacs]{revtex4}

\usepackage{graphics}
\usepackage{graphicx}
\usepackage{dcolumn}
\usepackage{bm}
\usepackage{amsmath,amssymb}

\newcommand{\eq}[2]{\begin{equation} \label{#1} #2 \end{equation}}

\begin{document}

\title{Talbot effect in binary waveguide arrays}
\author{Minh C. Tran $^{1,2,*}$ and Truong X. Tran$^{3,\dag}$}

\affiliation{$^{1}$ Atomic Molecular and Optical Physics Research Group, Science and Technology Advanced Institute, Van Lang University, Ho Chi Minh City, Vietnam\\
 $^{2}$ Faculty of Applied Technology, School of Technology, Van Lang University, Ho Chi Minh City, Vietnam\\
  $^{3}$ Department of Physics, Le Quy Don Technical University, 236 Hoang Quoc Viet street, 10000 Hanoi, Vietnam\\
 $^{*}$ trancongminh@vlu.edu.vn\\
 $\dag$ Corresponding author: tranxtr@gmail.com}
\date{\today}

\begin{abstract}
We study the Talbot effect in binary waveguide arrays (BWAs). Like in conventional waveguide arrays, the Talbot effect can only occur if the input signal has the period equal to $N$ = 1, 2, 3, 4, and 6 in the transverse direction. However, unlike in conventional waveguide arrays, for observation of the Talbot effect with $N$ = 3, 4, and 6 in BWAs, parameter $\sigma$ representing half of the propagation constant mismatch between two adjacent waveguides must have some specific values. Meanwhile, for observation of the Talbot effect with $N$ = 1 and 2 in BWAs, $\sigma$ can get any real values. We also analytically derive the Talbot distance along the longitudinal axis of BWAs where the recurrence of the input signal happens both in phase and intensity. Moreover, we also analytically find the intensity period where the field intensity is repeated during propagation. In some cases, the intensity period is equal to half of the Talbot distance, whereas in other cases, these two periods are just equal to each other. All these new analytical results are perfectly confirmed by beam propagation simulations in BWAs.
\end{abstract}
\maketitle

\section{I. INTRODUCTION}
\label{Introduction}

Waveguide arrays (WAs) are a remarkable platform possessing a rich variety of discrete photonic effects, for instance, the discrete diffraction \cite{jones}, discrete optical solitons \cite{christodoulides,kivshar,lederer,tranBBDS}, discrete Talbot effect \cite{talbot,iwanow}, diffractive resonant radiation \cite{tranresonant1}.

Remarkably, one can use WAs to simulate nonrelativistic quantum effects like Bloch oscillations \cite{bloch,blochZener,pertsch,lenz} and Zener tunneling \cite{ghulinyan,trompeter} because it is possible to convert the coupled-mode equations describing the propagation of light beams in WAs into the Schr\"{o}dinger equation. In particular, one can simulate basic relativistic quantum effects by optical analogues in binary waveguide arrays (BWAs) - a special class of WAs in which two different types of waveguides are periodically laid out adjacent to each  other. This is possible because one can convert the coupled-mode equations in BWAs into the Dirac equation. Thanks to that, in BWAs one can observe optical analogues of many relativistic quantum phenomena such as {\em Zitterbewegung} \cite{zitterbewegung}, Dirac solitons (DSs) \cite{trandirac2}, Klein tunneling \cite{klein,kleinLonghi,kleinDreisow,tranKT1,tranKT2}, Sauter effect \cite{sauter,kleinLonghi,tranSauter}, electron-positron pair production and annihilation \cite{longhiPP,dreisowPP,tranPP1,tranPP2}, and topological Jackiw-Rebbi states \cite{jackiw,tranjr4}.

The Talbot effect was first studied by Talbot in 1836 \cite{talbot}. This is the recurrence during propagation of any periodic one-dimensional optical inputs in continuous media at even integer multiples of the Talbot distance $z_{T} = d^{2}_{T}/\lambda$ where $d_{T}$ is the transverse period of the optical input field with wavelength $\lambda$ \cite{talbot,lederer,rayleigh}. The Talbot effect is a direct result of Fresnel diffraction \cite{winthrop}. Apart from integer Talbot effect, the so-called fractional Talbot effect can also exist when the recurrence of the optical input happens at distances that are rational multiples of $z_{T}$ such that $z/z_{T}$ = $p/q$ where $p$ and $q$ are relatively prime integers \cite{rayleigh,szwaykowski}. In addition, the resulting image is fractal if the ratio $z/z_{T}$ is irrational \cite{berry}. This kind of relationship between the Talbot effect and numbers can be used to factorize integers \cite{clauser}. In terms of applications, the Talbot effect can be exploited in the spatial domain to design the so-called Talbot-cavity semiconductor lasers \cite{mehuys}, or in the temporal domain to generate frequency combs \cite{azana}.

The Talbot effect belongs to a more general family of effects showing wave packet revivals \cite{robinett} which can exist in other areas of physics such as in Bose-Einstein condensate \cite{deng}, in atom optics \cite{chapman}, in atomic Rydberg electron wave packets \cite{eberly}, in quantum billiards and carpets  \cite{kaplan}, in Bloch oscillations \cite{bloch,blochZener,pertsch,lenz}, and Bloch-Zener oscillations \cite{breid,longhiZener,dreisowZener}. For the review of recent advances in the Talbot effect in classical optics, nonlinear optics and quantum optics, one can refer to Ref. \cite{wen}.

In discrete media such as conventional WAs consisting of identical waveguides, the Talbot effect was first studied in Ref. \cite{iwanow}. It turns out that in WAs, the Talbot effect can only happen when the transverse period $N$ of the optical inputs belongs to the set $N \in \{1, 2, 3, 4, 6\}$. This is totally different from the Talbot effect in continuous media where the revivals are independent of the input period.

In this work, inspired by the earlier achievements in investigating the Talbot effect in WAs, we investigate the Talbot effect in BWAs. We show that if the period $N$ of the input fields is 1 or 2, then the Talbot effect in BWAs can always happen in the linear regime, just like in WAs. However, if the period $N$ belongs to the set $N \in \{3, 4, 6\}$, then the Talbot effect in BWAs can only happen if the so-called Dirac mass $\sigma$ representing the binary nature of BWAs gets specific values which can be analytically calculated. This feature is totally different from the Talbot effect in WAs with the same period $N$. We also show that one can get exact analytical expressions for the Talbot distance $z_{T}$ and the intensity period $z_{I}$ where the recurrence of the input signals along the longitudinal axis of BWAs takes place. These analytical results are in perfect agreement with simulation results for the beam propagation in BWAs. The paper is organized as follows. In Section II we provide the theoretical background for the Talbot effect in BWAs. In Section III we verify the analytical results obtained in Section II by comparing them with simulation results. Finally, in Section IV we summarize our results and finish with concluding remarks.

\section{II. THEORY OF THE TALBOT EFFECT IN BINARY WAVEGUIDE ARRAYS}
\label{WA}

The beam light propagation in a BWA in the linear regime is governed by the following dimensionless coupled-mode equations \cite{sukhorukov1}:
\begin{equation} \label{CWCM} 
    i\frac{{da_n }}{{dz}} =  - \kappa \left[ {a_{n + 1}  + a_{n - 1} } \right] + \left( { - 1} \right)^n \sigma a_n,
\end{equation}
where $a_{n}$ represents the electric amplitude in the $n$th waveguide; $z$ denotes the variable along the longitudinal axis of BWAs; 2$\sigma$ and $\kappa$ represent, the propagation mismatch and the coupling coefficient between two neighboring waveguides of BWAs, respectively. One can always normalize variables in Eqs. (\ref{CWCM}) such that $\kappa$ is equal to unity. Note also that with any BWA one can always switch the positive and negative values of $\sigma$ just by shifting the waveguide position $n$ of the BWA by an \emph{odd} number. In certain conditions, one can convert Eqs. (\ref{CWCM}) into the Dirac equation in relativistic quantum mechanics \cite{zitterbewegung,tranSauter} in which parameter $\sigma$ plays the role of the Dirac mass. Thanks to that, one can use BWAs to simulate relativistic quantum effects as mentioned in Introduction.

By inserting a plane wave form
\eq{planewave}{a_{n}(Q) \sim \textrm{exp}{[i(Qn - \omega z)]},}
into Eqs. (\ref{CWCM}), one can get the following dispersion relationship \cite{sukhorukov1}:
\eq{dispersionBWA}{\omega_{\pm}(Q) =  \pm \sqrt{\sigma^{2} + 4\kappa^{2} \textrm{cos}^{2}Q},}
where the dimensionless quantity $Q$ is the normalized transverse wavenumber. We will refer to $Q$ just as the wavenumber in the rest of this work for the sake of brevity. This wavenumber is the phase difference between optical signals in two adjacent waveguides. Parameter $\omega$ in Eq. (\ref{planewave}) is the longitudinal wavenumber of the plane wave. Because around the Dirac points when $Q = \pm\pi/2$ two curves $\omega_{\pm}$ governed by Eqs. (\ref{dispersionBWA}) can be approximately reduced to the typical hyperbolas representing the dispersion curves of free electrons governed by the Dirac equation, it is the reason why one can use BWAs to simulate relativistic quantum effects. The dispersion relationship (\ref{dispersionBWA}) is used below to find conditions for the input period $N$ and calculate the Talbot distance $z_{T}$ and the intensity period $z_{I}$ of the Talbot effect in BWAs.

For conventional WAs consisting of just identical waveguides, i.e., when $\sigma$ = 0 in Eqs. (\ref{CWCM}), one can obtain the following dispersion relationship \cite{jones,lederer}:
\eq{dispersionWA}{\omega(Q) =  -2 \kappa \textrm{cos}Q.}
Note that because we use the sign (-) for $\omega z$ in Eq. (\ref{planewave}), we have the sign (-) on the right-hand side of Eq. (\ref{dispersionWA}). Otherwise, Eq. (\ref{dispersionWA}) will be the same as Eq. (2.5) in Ref. \cite{lederer}. The dispersion relationship (\ref{dispersionWA}) has been used in Ref. \cite{iwanow} to find conditions for the input period $N$ and calculate the Talbot distance in conventional WAs.

As reasoned in Ref. \cite{iwanow}, for the Talbot effect to take place in BWAs, the input field distribution should be periodic with the period $N$, i.e., $a_{n+N} = a_{n}$. Then from Eq. (\ref{planewave}), we have $\textrm{exp}{(iQN)}$ = 1, or the wavenumber $Q$ can get following eigenvalues:
\eq{Qm}{Q_{m} =  m\frac{2 \pi}{N},}
with $m$ belonging to the set $m \in \{0, 1, 2, ... N-1\}$. From these eigenvalues for $Q_{m}$, one can straightforwardly calculate the eigenvalues for $\omega_{m}$ by using dispersion relationship in the form of Eqs. (\ref{dispersionBWA}):

\eq{eigenvalueomega}{\omega_{m} =  \sqrt{\sigma^{2} + 4\kappa^{2} \textrm{cos}^{2}\left(m\frac{2 \pi}{N}\right)},}
where we just keep the plus sign in front of the square root to simplify the reasoning below. This simplification does not lose any generality in considering the requirement for input period $N$, calculating the Talbot distance $z_{T}$ and the intensity period $z_{I}$ below in this work.

In general, due to the periodicity, the field evolution at the $n$th waveguide can be described by the orthonormal set of functions \cite{iwanow}:
\eq{set}{u_{n}^{(m)} = N^{-1/2}\textrm{exp}{(inQ_{m})}\textrm{exp}{(-i\omega_{m} z)}.}
Therefore, the field $a_{n}$ with the input period $N$ can be written as the following linear combination of the orthonormal functions \cite{iwanow}:
\eq{superposition}{a_{n}^{(N)} = \sum_{m=0}^{N-1} c_{m}u_{n}^{(m)}.}

Consequently, it is clear from Eq. (\ref{set}) and Eq. (\ref{superposition}) that the field recurrence both in phase and intensity is possible at the Talbot distance $z_{T}$ only if:
\eq{omegaZt}{\omega_{m} z_{T} = 2\nu\pi,}
where $\nu$ is an integer. Therefore, the ratio of any two eigenvalues $\omega_{m}$ must be a rational number. Specifically, we need to have:
\eq{ratio}{\frac{\omega_{m}}{\omega_{h}} = \frac{p}{q},}
where $p$ and $q$ are relatively prime integers, and subscript $h$ belongs to the set $h \in \{0, 1, 2, ... N-1\}$ just like subscript $m$. Inserting Eq. (\ref{eigenvalueomega}) into Eq. (\ref{ratio}), we have the following condition for the ratio $\omega_{1}/\omega_{0}$ (with the period $N > 1$) for observing the Talbot effect in BWAs:
\eq{ratioDispersion}{\frac{\omega_{1}}{\omega_{0}}= \frac{\sqrt{\sigma^{2} + 4\kappa^{2} \textrm{cos}^{2}(\frac{2 \pi}{N})}}{\sqrt{\sigma^{2} + 4\kappa^{2}}} = \frac{p}{q}.}

If the input period $N$ = 1, from Eq. (\ref{Qm}) one can see that there is just one value for $m = 0$, and the wavenumber $Q_{0}$ = 0 and  $\omega_{0} =  \sqrt{\sigma^2 + 4\kappa^2}$ as seen from Eq. (\ref{eigenvalueomega}). As a result, from Eq. (\ref{omegaZt}) one can easily get the formula for the Talbot distance in BWAs when the input period $N = 1$ as follows:
\eq{TalbotdistanceN1}{z_{T} = \frac{2\pi}{\sqrt{\sigma^2 + 4\kappa^2}}.}

This Talbot distance for $N = 1$ is always satisfied with any real value of $\sigma$, just like what happens with $N = 2$, but unlike what happens with other values of $N$ as shown below.

Note that the Talbot distance given by Eq. (\ref{TalbotdistanceN1}) for $N$ = 1 is the period along the longitudinal $z$-axis where both the phase and intensity of the input pattern are repeated. If one is just interested in the recurrence of the absolute values $|a_{n}|$, then the intensity period $z_{I}$ is half of the Talbot distance given by Eq. (\ref{TalbotdistanceN1}), because in this case it is just necessary to require that $\omega_{0}z_{I} = \pi$ instead of $2\pi$ as in Eq. (\ref{omegaZt}), therefore, we have:
\eq{intensityperiodN1}{z_{I} = \frac{\pi}{\sqrt{\sigma^2 + 4\kappa^2}} = \frac{z_{T}}{2}.}
Indeed, at the distance $z_{I}$, all the orthonormal functions $u^{(m)}_{n}(z=z_{I})$ in Eq. (\ref{set}) will be out-of-phase with their corresponding input components $u^{(m)}_{n}(z = 0)$, therefore, all the field components $a^{(N)}_{n}(z=z_{I})$ in Eq. (\ref{superposition}) will be out-of-phase with their input field components $a^{(N)}_{n}(z=0)$. For instance, if the input pattern is $a^{(N=1)}(z=0) = \{1, 1, 1, ...\}$, then at the distance $z_{I}$ the field components will be $a^{(N=1)}(z=z_{I}) = \{-1, -1, -1, ...\}$, and at the Talbot distance $z_{T} = 2z_{I}$ we will have the field components $a^{(N=1)}(z=z_{T}) = \{1, 1, 1, ...\}$. As a result, the absolute values of the field $|a_{n}|$ will evolve with the intensity period $z_{I} = 0.5 z_{T}$ as given by Eq. (\ref{TalbotdistanceN1}) for $N$ = 1.

If the input period $N = 2$, as seen from Eq. (\ref{Qm}), $m$ can only take two values 0 and 1. In this case, according to Eq. (\ref{Qm}), $Q_{0}$ = 0 and $Q_{1} = \pi$. As a result, both eigenvalues $\omega_{0}$ and $\omega_{1}$ are equal to each other and get the value $\sqrt{\sigma^2 + 4\kappa^2}$ as seen from Eq. (\ref{eigenvalueomega}). Consequently, when the input period $N = 2$, just like in the case with $N = 1$, the Talbot distance is also given by Eq. (\ref{TalbotdistanceN1}), and $\sigma$ can get any real values. Because the only two eigenvalues $\omega_{0}$ and $\omega_{1}$ are equal to each other when $N$ = 2, the intensity period in this case is also calculated from the condition $\omega_{0}z_{I} = \omega_{1}z_{I} = \pi$ and given by Eq. (\ref{intensityperiodN1}), just as in the case with $N = 1$ explained above.

If the input period $N > 2$, the situation is totally different from what happens with $N$ = 1 and 2. Indeed, from the requirement in the form of Eq. (\ref{ratioDispersion}), Dirac mass $\sigma$ must now get following specific values in order for the Talbot effect to take place in BWAs:

\eq{diracmass}{\sigma = \pm2\kappa\frac{\sqrt{p^2 - q^2 \textrm{cos}^{2}(\frac{2 \pi}{N})}} {\sqrt{q^2 - p^2}}.}
Note that we can only use relatively prime integers $p$ and $q$ such that $\sigma$ calculated from Eq. (\ref{diracmass}) is real.

If the input period $N = 3$, then $m$ can get three values: $m$ = 0, 1, and 2. From Eq. (\ref{eigenvalueomega}), one can see that $\omega_{1}$ = $\omega_{2}$ = $\sqrt{\sigma^2 + \kappa^2}$, whereas $\omega_{0} = \sqrt{\sigma^2 + 4\kappa^2}$. As a result, the Talbot distance in BWAs with $N = 3$ can be calculated by using Eq. (\ref{omegaZt}) with $m$ = 0 and $\nu = q$, or $m$ = 1 (or 2) and $\nu = p$. Namely, the Talbot distance in BWAs with $N = 3$ is given as follows:

\eq{TalbotdistanceN3}{z_{T} = \frac{2\pi q}{\sqrt{\sigma^2 + 4\kappa^2}}.}

If the input period $N = 4$, then $m$ can get four values: $m$ = 0, 1, 2, and 3. From Eq. (\ref{eigenvalueomega}), one can see that $\omega_{1}$ = $\omega_{2}$ = $\omega_{3}$ = $|\sigma|$, whereas $\omega_{0} = \sqrt{\sigma^2 + 4\kappa^2}$. As a result, the Talbot distance in BWAs with $N = 4$ can be calculated by using Eq. (\ref{omegaZt}) with $m$ = 0 and $\nu = q$, or $m$ = 1 (or 2, or 3) and $\nu = p$. Consequently, the Talbot distance in BWAs with $N = 4$ is also given by Eq. (\ref{TalbotdistanceN3}), just the same form as with $N = 3$.

Similarly, if the input period $N = 6$, then $m$ can get six values: $m$ = 0, 1, 2, 3, 4, and 5. From Eq. (\ref{eigenvalueomega}), one can see that $\omega_{m}$ = $\sqrt{\sigma^2 + \kappa^2}$ with $m$ = 1, 2, 4, and 5, whereas $\omega_{0} = \omega_{3} = \sqrt{\sigma^2 + 4\kappa^2}$. As a result, the Talbot distance in BWAs with $N = 6$ can be calculated by using Eq. (\ref{omegaZt}) with $m$ = 0 (or 3) and $\nu = q$, or $m$ = 1 (or 2, 4, 5) and $\nu = p$. Consequently, the Talbot distance in BWAs with $N = 6$ is also given by Eq. (\ref{TalbotdistanceN3}), just the same form as with $N = 3$ and 4.

However, the situation will be different with other values of the input period $N$, i.e., when $N$ = 5 and $\geq 7$. In this case, due to the presence of $\textrm{cos}(2\pi m/N)$ in Eq. (\ref{eigenvalueomega}), there are at least \emph{three} different values for all possible eigenvalues $\omega_{m}$. As a result, it is extremely difficult to simultaneously satisfy two conditions similar to Eq. (\ref{ratio}) where the ratio of all possible two eigenvalues $\omega_{m}$ must be rational numbers. For instance, with $N$ = 5, at first let us set $q$ = 2 and $p$ = 1 as an example, then we have $\sigma$ = 0.9078 as given by Eq. (\ref{diracmass}), after that, from Eq. (\ref{eigenvalueomega}) we have $\omega_{0}$ = 2.1964, $\omega_{1} = \omega_{4}$ = 1.0982, and $\omega_{2} = \omega_{3} = 1.8553$. Therefore, from Eq. (\ref{omegaZt}) one can see that the field recurrence at the distance $z_{T}$ is only possible if one can find three integers $f_{0}$, $f_{1}$ and $f_{2}$ such that $\omega_{0}$ : $\omega_{1}$ : $\omega_{2}$ = $f_{0}$ : $f_{1}$ : $f_{2}$. Consequently, one can choose $f_{0}$ = 21964, $f_{1}$ = 10982, and $f_{2}$ = 18553 as the set of minimum values. In this case, $z_{T}$ = 20000$\pi$ as given by Eq. (\ref{omegaZt}). This distance is extremely large for both simulations and for real BWAs. In addition, for such a long distance, we need to use BWAs with a huge total number of waveguides (so that the edge effect does not have any significant influence on the field evolution at the central part of BWAs where we observe the Talbot effect). Therefore, theoretically, it is possible to observe the Talbot effect in BWAs with $N$ = 5 and $\geq 7$, but practically, it is extremely hard to do that in experiments and simulations.

Note that when $N$ = 1 and 2, we have all eigenvalues $\omega_{m}$ equal to each other in each case, therefore, we can easily get the intensity period $z_{I} = 0.5z_{T}$ where $\omega_{m}z_{I}$ = $\pi$ with all possible values of $m$ for each case. On the contrary, with $N$ = 3, 4, and 6, all eigenvalues $\omega_{m}$ are not equal to each other (there are \emph{two} different values for all of them). Therefore, the field recurrence in intensity at half of the Talbot distance can only happen if both two relatively prime integers $p$ and $q$ in Eq. (\ref{ratioDispersion}) are \emph{odd}. For instance, with $N$ = 3 at the Talbot distance $z_{T}$ we have $\omega_{0}z_{T} = 2q\pi$ and $\omega_{1}z_{T} = \omega_{2}z_{T} = 2p\pi$. Therefore, at half of the Talbot distance, i.e., at the intensity period $z_{I}$ given below:
\eq{intensityperiodN3}{z_{I} = \frac{\pi q}{\sqrt{\sigma^2 + 4\kappa^2}}}
we have $\omega_{0}z_{I} = q\pi$ and $\omega_{1}z_{I} = \omega_{2}z_{I} = p\pi$. And because two relatively prime integers $p$ and $q$ are both odd numbers in this case, at the distance $z_{I}$ all the field components are out-of-phase with the input components. As a result, the intensity $|a_{n}|$ is repeated at the multiples of this intensity period $z_{I}$ for the same reasoning when $N$ = 1 explained below Eq. (\ref{intensityperiodN1}). This feature is perfectly confirmed by simulations in Section III.

On the contrary, if one of the two numbers $p$ and $q$ is odd and the other is even, then at the distance $z_{I}$ some orthonormal functions $u^{(m)}_{n}$ given by Eqs. (\ref{set}) is out-of-phase, but the others are in-phase with their corresponding orthonormal functions at the input. Therefore, in this case, at the distance $z_{I}$ given by Eq. (\ref{intensityperiodN3}), the field components $a^{(N)}_{n}$ given by Eq. (\ref{superposition}) are not always out-of-phase with their corresponding input field components. As a result, in this case, we have just one longitudinal period for the field evolution which is the Talbot distance $z_{T}$ given by Eq. (\ref{TalbotdistanceN3}), and which is also the intensity period $z_{I}$ at the same time , i.e., $z_{I}$ = $z_{T}$. This feature is also perfectly confirmed by simulations in Section III.

In short, like in WAs, one can observe the Talbot effect if the input period $N$ belongs to the set $N \in \{1, 2, 3, 4, 6\}$. However, unlike in WAs, for observation of the Talbot effect in BWAs with $N$ = 3, 4, and 6, the Dirac mass $\sigma$ must get some specific values, whereas with $N$ = 1 and 2 this requirement is lifted. The Talbot distance given by Eq. (\ref{TalbotdistanceN1}) and Eq. (\ref{TalbotdistanceN3}) is the period along the $z$-axis when both the intensity and phase of the input pattern are repeated. If $N$ = 1 and 2, one can always find the intensity period $z_{I}$ (half of the Talbot distance) in the form of Eq. (\ref{intensityperiodN1}) where the intensity $|a_{n}|$ is repeated. However, if $N$ = 3, 4, and 6, the intensity period $z_{I}$ given by Eq. (\ref{intensityperiodN3}) (half of the Talbot distance) is valid if both two relatively prime integers $p$ and $q$ in Eq. (\ref{ratioDispersion}) are odd numbers. Otherwise, the intensity period $z_{I}$ is exactly equal to the Talbot distance $z_{T}$ given by Eq. (\ref{TalbotdistanceN3}) when $N$ = 3, 4, and 6. All the analytical results obtained in this Section for the Talbot effect in BWAs are in perfect agreement with the direct simulations of the beam propagation in BWAs in the linear regime in Section III.

\section{III. VERIFICATION OF THEORETICAL RESULTS WITH SIMULATIONS}
\label{LBWA}

\begin{figure}[htb]
  \centering \includegraphics[width=0.45\textwidth]{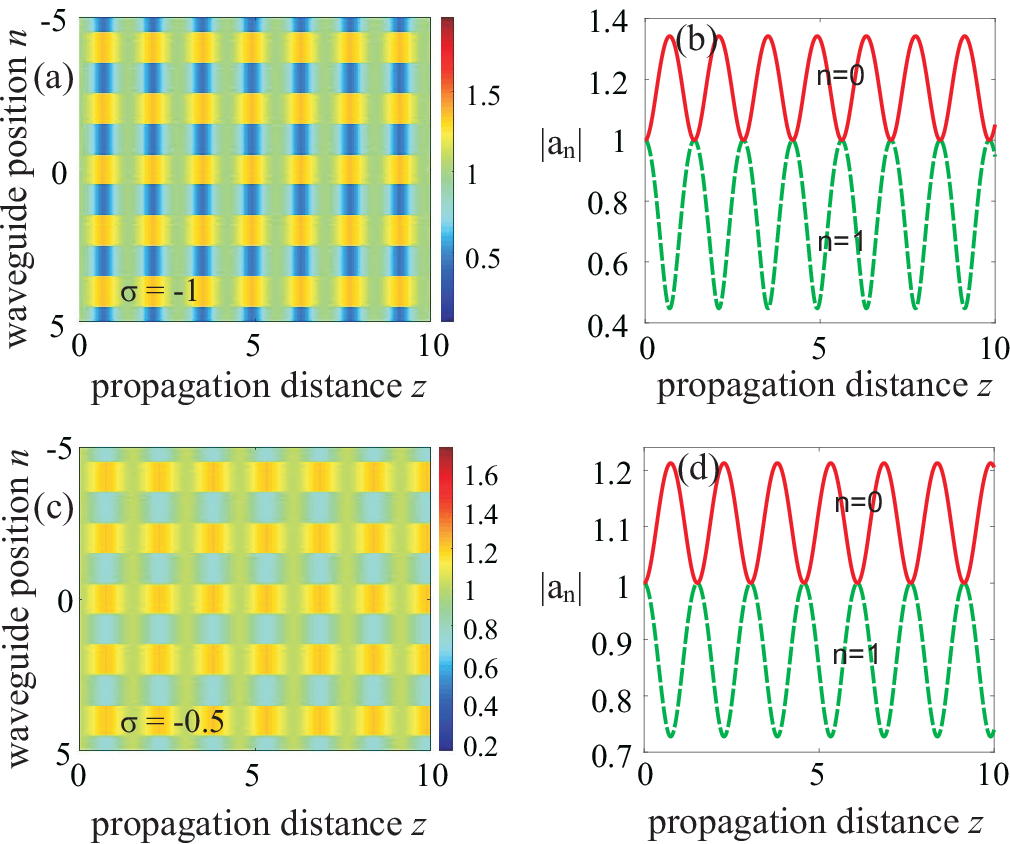}
  \caption{\small{(Color online) Talbot effect in BWAs with $N$ = 1. (a) Propagation of beams with the transverse input period $N = 1$ and the input pattern \{1,1,1,1,...\} when $\sigma = -1$. (b) Evolution of $|a_{n}|$ along the $z$-axis taken from (a) when $n$ = 0 and $n$ = 1. (b,c) The same as (a,b), but when $\sigma = -0.5$.}}
  \label{fig1}
\end{figure}

Now it is time for us to simulate the Talbot effect in BWAs and verify the analytical results obtained in Section II. We first start with the input period $N$ = 1, i.e., when the input pattern is $a_{n} = \{1, 1, 1, 1,...\}$ as the initial condition to numerically integrate Eqs. (\ref{CWCM}) by using the fourth-order Runge-Kutta algorithm. In Fig. \ref{fig1}(a) we show the evolution of the absolute values $|a_{n}|$ in the $(n,z)$-plane with $\sigma = -1$. Note that the coupling coefficient is fixed at $\kappa$ = 1 and the total number of waveguides in BWAs is 201 throughout this work.

In Fig. \ref{fig1}(b) we plot the evolution of $|a_{n}|$ at the two central waveguides with $n = 0$ (upper solid red curve) and $n = 1$ (lower dashed green curve). Unlike the situation with $N$ = 1 in WAs where the pattern does not change at all during propagation (i.e., $|a_{n}|$ does not depend on $z$), in BWAs the input field pattern does change during propagation, but periodically repeats with the Talbot distance $z_{T}$ = 2.8099 as predicted by Eq. (\ref{TalbotdistanceN1}). Actually, the period for the recurrence of the absolute values $|a_{n}|$ when the input period $N$ = 1 is $z_{I}$ = 1.4049 (half of the Talbot distance) as given by Eq. (\ref{intensityperiodN1}) and confirmed by Figs. \ref{fig1}(a,b). Our phase analysis (not included here) by using function \emph{angle} in Matlab for complex values $a_{n}$ at specific propagation distances $z$ confirms that at the Talbot distance and its multiples, the phase of $a_{n}$ is a multiple of $2\pi$, whereas its phase at the intensity period $z_{I}$ given by Eq. (\ref{intensityperiodN1}) is just $\pi$, i.e., out-of-phase with the input field, just as predicted in Section II.

In Figs. \ref{fig1}(c,d) we show the two panels just like in Figs. \ref{fig1}(a,b), but now $\sigma = -0.5$. As predicted by Eq. (\ref{TalbotdistanceN1}) and Eq. (\ref{intensityperiodN1}), the Talbot distance now is $z_{T} = 3.0478$ and the intensity period now is $z_{I}$ = 1.5239 as perfectly confirmed by Figs. \ref{fig1}(c,d). Our simulations (not included here) also confirm that the Talbot effect always take place in BWAs with the input period $N$ = 1 with any real value of $\sigma$ in use.

\begin{figure}[htb]
  \centering \includegraphics[width=0.45\textwidth]{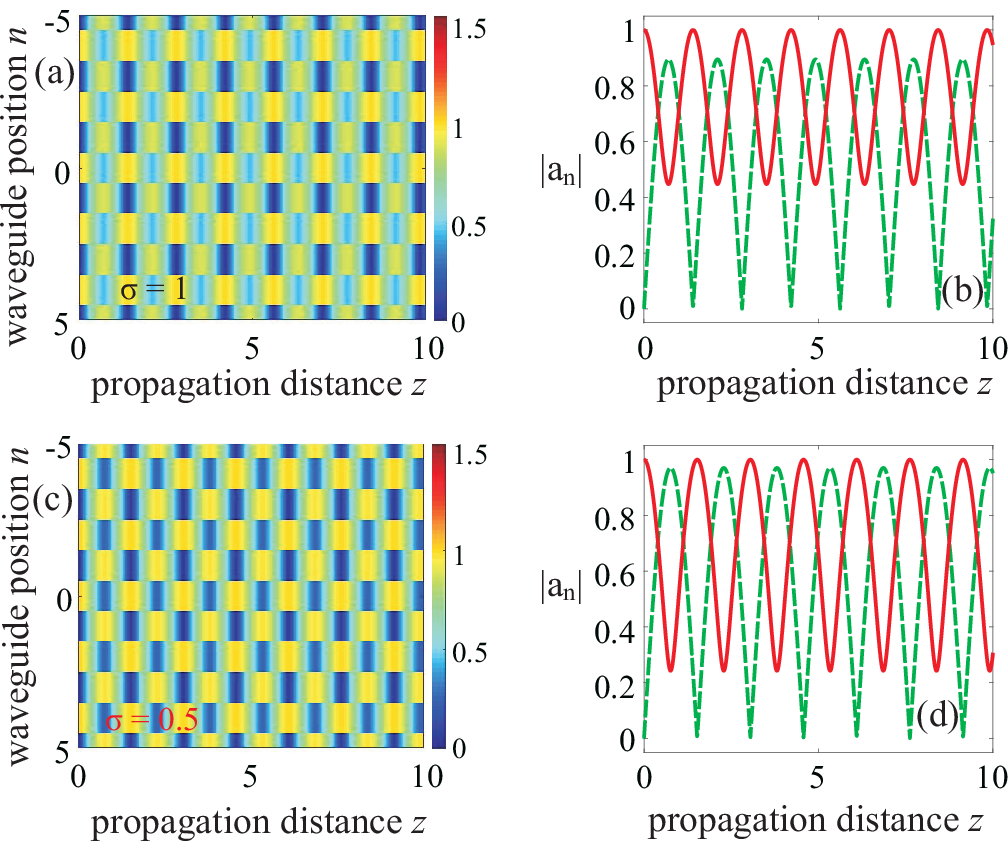}
  \caption{\small{(Color online) The same as Fig. \ref{fig1}, but now with the transverse input period $N = 2$ and the input pattern \{1,0,1,0,...\}. Other parameters: $\sigma$ = 1 in (a,b) and $\sigma$ = 0.5 in (c,d).}}
  \label{fig2}
\end{figure}

The simulation for the Talbot effect in BWAs when $N$ = 2 with the input field $a_{n} = \{1,0,1,0,...\}$ is shown in Fig. \ref{fig2}. All four panels of Fig. \ref{fig2} just show the same quantities as corresponding panels in Fig. \ref{fig1}. We set $\sigma$ = 1 and 0.5 in Figs. \ref{fig2}(a,b) and Figs. \ref{fig2}(c,d), respectively. For $N$ = 2, the Talbot distance $z_{T}$ and the intensity period $z_{I}$ are also given by Eq. (\ref{TalbotdistanceN1}) and Eq. (\ref{intensityperiodN1}), respectively, just like when $N$ = 1. As one can clearly see, the intensity period in Figs. \ref{fig2}(a,b) is $z_{I}$ = 1.4049 as in Figs. \ref{fig1}(a,b); and the intensity period in Figs. \ref{fig2}(c,d) is $z_{I}$ = 1.5239 as in Figs. \ref{fig1}(c,d), as perfectly predicted by Eq. (\ref{intensityperiodN1}).

\begin{figure}[htb]
  \centering \includegraphics[width=0.45\textwidth]{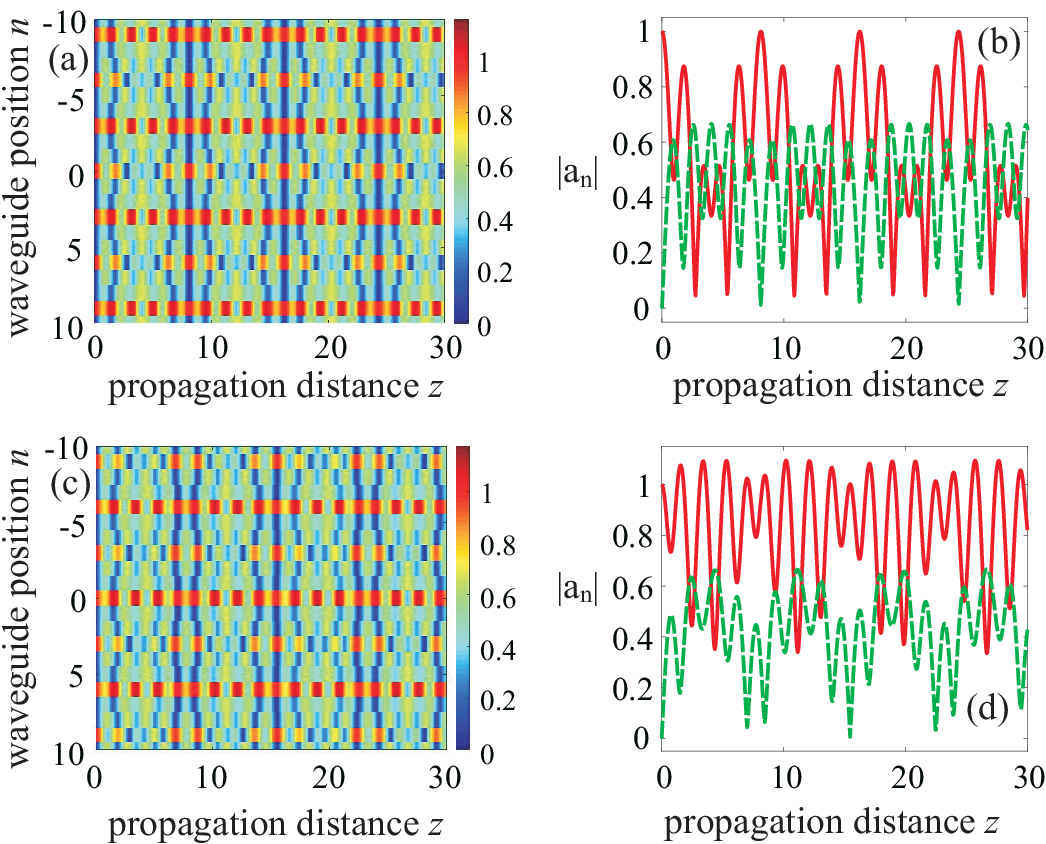}
  \caption{\small{(Color online) (a) Propagation of beams with the transverse input period $N = 3$ and the input pattern \{1,0,0,1,0,0,1,...\} when $\sigma$ is calculated from Eq. (\ref{diracmass}) with $p$ = 2 and $q$ = 3. (b) Evolution of $|a_{n}|$ along the $z$-axis taken from (a) when $n$ = 0 and $n$ = 1. The same as (a,b), but when $\sigma = 1$.}}
  \label{fig3}
\end{figure}

Unlike the case with $N$ = 1 and 2 when there is no requirement for $\sigma$ as long as this parameter is real, now we show the simulation results of the Talbot effect with $N > 2$ when $\sigma$ must get the specific values given by Eq. (\ref{diracmass}) if we want to observe the Talbot effect in BWAs. We first start with $N$ = 3 in Fig. \ref{fig3} where the input pattern $a_{n} = \{1,0,0,1,0,0,1,...\}$ is used. In Fig. \ref{fig3}(a) we show the Talbot effect when $\sigma$ is set at the value $\sigma$ = 1.1832 given by Eq. (\ref{diracmass}) with two relatively prime integers $p$ = 2 and $q$ = 3. As shown in Fig. \ref{fig3}(a), the evolution of $|a_{n}|$ is periodically with the intensity period $z_{I}$ equal to the Talbot distance $z_{T}$ = 8.1116 as given by Eq. (\ref{TalbotdistanceN3}). This periodicity is also confirmed by Fig. \ref{fig3}(b) where we show the evolution of $|a_{n}|$ in the two central waveguides with $n$ = 0 (solid red curve) and $n$ = 1 (dashed green curve). Our phase analysis (not included here) also confirms that at the Talbot distance $z_{T}$ = 8.1116, the phase of all the components $a_{n}$ is equal to $2\pi$, as expected. As explained in Section II, when $N$ = 3 and larger, if one of the two relatively prime numbers $p$ and $q$ in Eq. (\ref{ratioDispersion}) and Eq. (\ref{diracmass}) is even (2) while the other is odd (3), then the intensity period $z_{I}$ is the same as the Talbot distance $z_{T}$ where both the intensity and phase of the field are repeated, unlike the situation with $N$ = 1 and 2 when there is always the intensity period $z_{I} = 0.5 z_{T}$ where the intensity $|a_{n}|$ of the field is repeated.

In Figs. \ref{fig3}(c,d) we show the same quantities as in Figs. \ref{fig3}(a,b), respectively, but now with $\sigma$ = 1 instead of the required value $\sigma$ = 1.1832 given by Eq. (\ref{diracmass}). As clearly seen in Figs. \ref{fig3}(c,d), the Talbot effect with strict periodicity does not exist in this case, just as explained in Section II.

\begin{figure}[htb]
  \centering \includegraphics[width=0.45\textwidth]{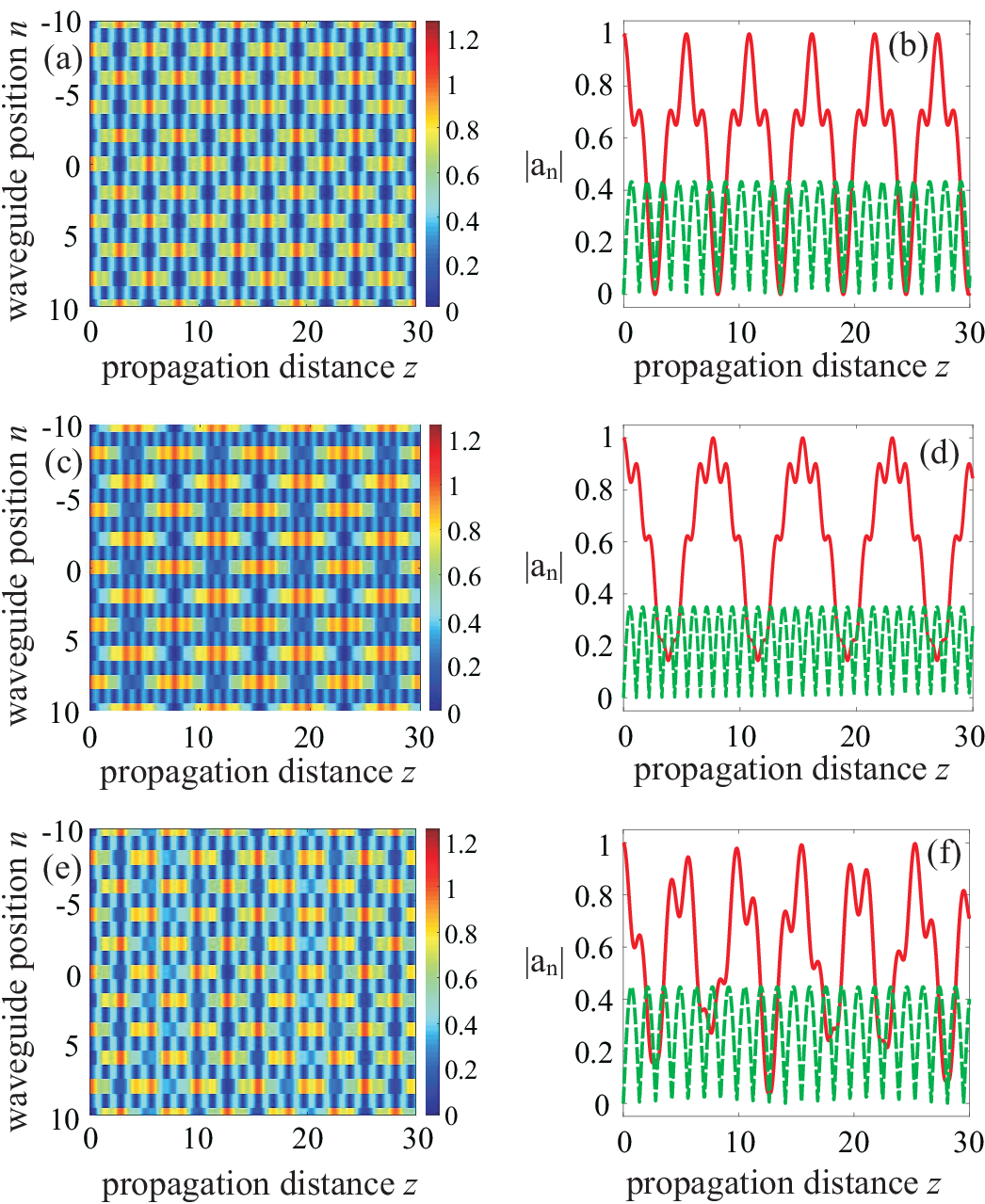}
  \caption{\small{(Color online) (a) Propagation of beams with the transverse input period $N = 4$ and the input pattern \{1,0,0,0,1,0,0,0,1,...\} when $\sigma$ is calculated from Eq. (\ref{diracmass}) with $p$ = 1 and $q$ = 2. (b) Evolution of $|a_{n}|$ along the $z$-axis taken from (a) when $n$ = 0 and $n$ = 1. (c,d) The same as (a,b), but when $\sigma$ is calculated from Eq. (\ref{diracmass}) with $p$ = 5 and $q$ = 7. (e,f) The same as (a,b), but when $\sigma = 1$.}}
  \label{fig4}
\end{figure}

In Fig. \ref{fig4} we investigate the beam evolution in BWAs with $N$ = 4 with the input pattern $a_{n} = \{1,0,0,0,1,0,0,0,1,...\}$. In Fig. \ref{fig4}(a) we show the Talbot effect when $\sigma$ is set at the value $\sigma$ = 1.1547 given by Eq. (\ref{diracmass}) with two relatively prime integers $p$ = 1 and $q$ = 2. As shown in Fig. \ref{fig4}(a), the evolution of $|a_{n}|$ is periodically with the intensity period equal to the Talbot distance $z_{T}$ = 5.4414 as given by Eq. (\ref{TalbotdistanceN3}). This periodicity is also clearly confirmed by Fig. \ref{fig4}(b) where we show the evolution of $|a_{n}|$ in the two central waveguides with $n$ = 0 (solid red curve) and $n$ = 1 (dashed green curve). Note again that because one of the two relatively prime numbers $p$ and $q$ in Eq. (\ref{ratioDispersion}) and Eq. (\ref{diracmass}) is even (2) while the other is odd (1), the intensity period $z_{T}$ must be the same as the Talbot distance $z_{T}$ as in Figs. \ref{fig3}(a,b), just as expected.

In Figs. \ref{fig4}(c,d) we show the same quantities as in Figs. \ref{fig4}(a,b), respectively, but now with $\sigma$ = 2.0412 given by Eq. (\ref{diracmass})
with two relatively prime integers $p$ = 5 and $q$ = 7. As predicted in Section II, when both $p$ and $q$ are \emph{odd} numbers for $N$ = 3, 4, and 6, then the field evolution must have two longitudinal periods: the Talbot distance $z_{T}$ given by Eq. (\ref{TalbotdistanceN3}) where both the field intensity and phase are repeated, and the intensity period $z_{I}$ (half of $z_{T}$) given by Eq. (\ref{intensityperiodN3}) where the field intensity is repeated. This is perfectly confirmed in Figs. \ref{fig4}(c,d) where the intensity period $z_{I}$ = 7.6953 (half of $z_{T}$) as exactly given by Eq. (\ref{intensityperiodN3}). Our phase analysis (not included here) also confirms that at the intensity period $z_{I} = 0.5 z_{T}$ the field in Figs. \ref{fig4}(c,d) is out-of-phase with the input field, whereas at the Talbot distance $z_{T}$ the field is in-phase with the input field, as exactly predicted in Section II.

In Figs. \ref{fig4}(e,f) we show the same quantities as in Figs. \ref{fig4}(a,b), respectively, but now with $\sigma$ = 1 instead of the required value $\sigma$ = 1.1547 given by Eq. (\ref{diracmass}) with $p$ = 1 and $q$ = 2. As clearly seen in Figs. \ref{fig4}(e,f), the Talbot effect with strict periodicity does not exist in this case, just as predicted in Section II.

\begin{figure}[htb]
  \centering \includegraphics[width=0.45\textwidth]{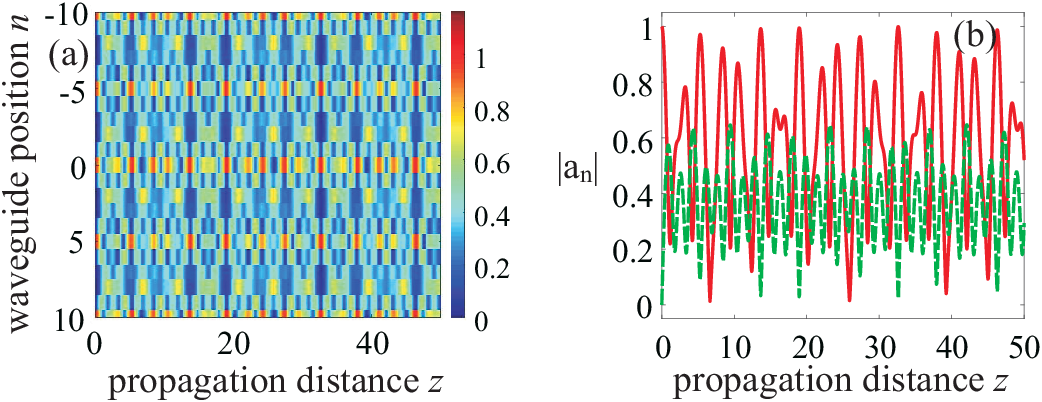}
  \caption{\small{(Color online) (a) Propagation of beams with the transverse input period $N = 5$ and the input pattern \{1,0,0,0,0,1,0,0,0,0,1,...\} when $\sigma$ is calculated from Eq. (\ref{diracmass}) with $p$ = 1 and $q$ = 3. (b) Evolution of $|a_{n}|$ along the $z$-axis taken from (a) when $n$ = 0 and $n$ = 1.}}
  \label{fig5}
\end{figure}

In Fig. \ref{fig5} we show the field evolution in BWAs when we use the input pattern $a_{n} = \{1,0,0,0,0,1,0,0,0,0,1,...\}$, i.e., with $N$ = 5. Although we set $\sigma$ = 0.2651 as given by Eq. (\ref{diracmass}) with $p$ = 1 and $q$ = 3, the Talbot effect cannot exist with $N$ = 5 as explained in Section II and similar to the case with WAs reported in Ref. \cite{iwanow}.

\begin{figure}[htb]
  \centering \includegraphics[width=0.45\textwidth]{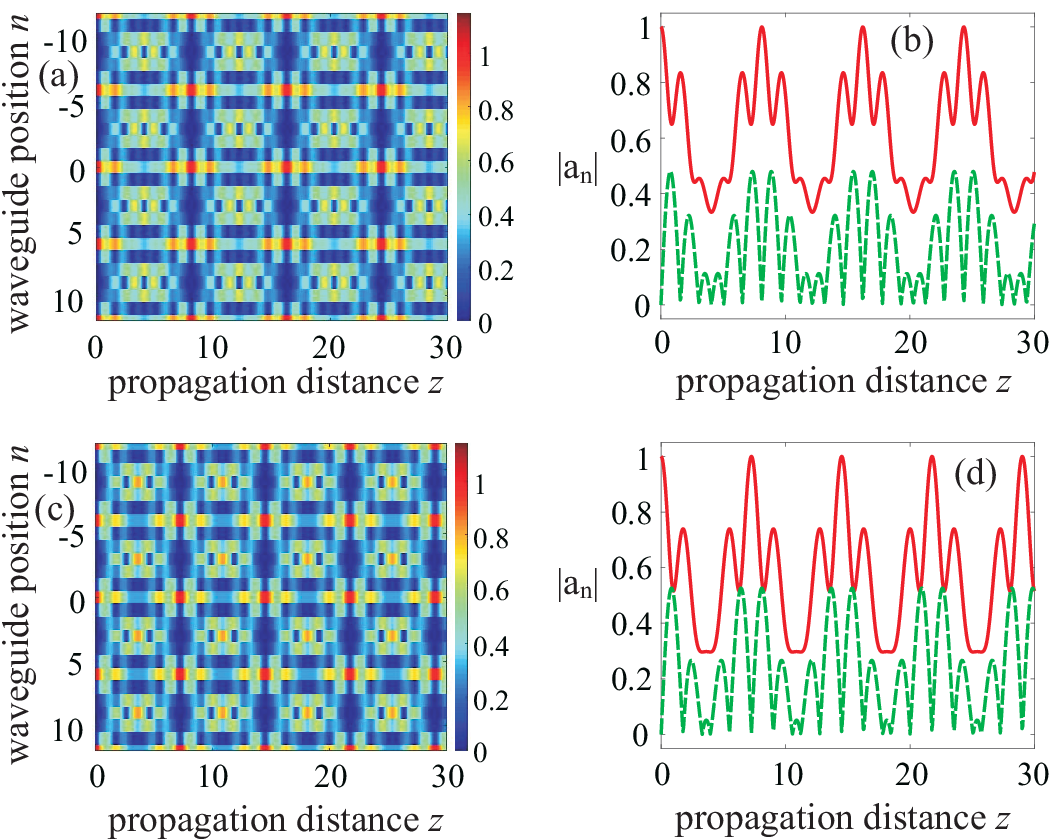}
  \caption{\small{(Color online) (a) Propagation of beams with the transverse input period $N = 6$ and the input pattern \{1,0,0,0,0,0,1,0,0,0,0,0,1,...\} when $\sigma$ is calculated from Eq. (\ref{diracmass}) with $p$ = 2 and $q$ = 3. (b) Evolution of $|a_{n}|$ along the $z$-axis taken from (a) when $n$ = 0 and $n$ = 1. (c,d) The same as (a,b), but when $p$ = 3 and $q$ = 5.}}
  \label{fig6}
\end{figure}

In Fig. \ref{fig6} we investigate the Talbot effect in BWAs with $N$ = 6 with the input pattern $a_{n} = \{1,0,0,0,0,0,1,0,0,0,0,0,1,...\}$. In Figs. \ref{fig6}(a,b) we set $\sigma$ = 1.1832 as given by Eq. (\ref{diracmass}) with two relatively prime integers $p$ = 2 and $q$ = 3. As shown in Fig. \ref{fig6}(a), the evolution of $|a_{n}|$ is perfectly periodic with the intensity period equal to the Talbot distance $z_{T}$ = 8.1116 as given by Eq. (\ref{TalbotdistanceN3}). This periodicity is also clearly confirmed by Fig. \ref{fig6}(b) where we show the evolution of $|a_{n}|$ in the two central waveguides with $n$ = 0 (solid red curve) and $n$ = 1 (dashed green curve).

The fact that the intensity period is exactly equal to the Talbot distance $z_{T}$ in Figs. \ref{fig6}(a,b) [and in Figs. \ref{fig4}(a,b)] has been explained in detail in Section II because in this case one of the two relatively prime integers is even ($p$ = 2) and the other is odd ($q$ = 3). This is totally different to the case where both $p$ and $q$ are odd as shown in Figs. \ref{fig6}(c,d) [and in Figs. \ref{fig4}(c,d) as well] where we predict in Section II that the intensity period $z_{I}$ must be equal to half of the Talbot distance. In Figs. \ref{fig6}(c,d) we show the same quantities as in Figs. \ref{fig6}(a,b), respectively, but now with $\sigma$ = 0.8292 as given by Eq. (\ref{diracmass}) with two relatively prime integers $p$ = 3 and $q$ = 5. As predicted by Eq. (\ref{TalbotdistanceN3}) and Eq. (\ref{intensityperiodN3}), the Talbot distance in this case is $z_{T}$ = 14.5104 and the intensity period is $z_{I}$ = 7.2552 (half of $z_{T}$) which is perfectly confirmed in Figs. \ref{fig6}(c,d).

\section{IV. CONCLUSIONS}
\label{conclusions}

We have systematically investigated the Talbot effect in BWAs in the linear regime. We have proved that the Talbot effect in BWAs is possible if $N \in \{1, 2, 3, 4, 6\}$, just like in WAs. However, if $N$ = 3, 4, and 6, then the Dirac mass $\sigma$ must take specific values depending on two relatively prime numbers $p$ and $q$ in order for the Talbot effect to take place, whereas there is no requirement for $\sigma$ at all (as long as it is real) if $N$ = 1 and 2. We have also analytically found the Talbot distance $z_{T}$ where the field intensity and phase are both repeated during propagation. Moreover, the intensity period $z_{I}$ where the intensity of the field is repeated during propagation has also been analytically found. If $N$ = 1 and 2, then the intensity period is always equal to half of the Talbot distance $z_{T}$. When $N$ = 3, 4, and 6, then the intensity period is exactly equal to half of the Talbot distance if two relatively prime numbers $p$ and $q$ constructing $\sigma$ are both odd numbers. In these cases, at the intensity period $z_{I}$ the field is out-of-phase with the input field. However, when one of those two numbers $p$ or $q$ is odd and the other is even, then the intensity period is exactly equal to the Talbot distance when $N$ = 3, 4, and 6. If $N$ is larger than 6, then it is practically impossible to observe the Talbot effect in BWAs. All these new analytical findings are in perfect agrement with the simulations for the beam propagation.


\begin{thebibliography}{99}

\bibitem{jones} A. L. Jones, J. Opt. Soc. Am. {\bf 55}, 261 (1965).
\bibitem{christodoulides} D. N. Christodoulides and R. I. Joseph, Opt. Lett. {\bf 13}, 794 (1988).
\bibitem{kivshar} Y. S. Kivshar and G. P. Agrawal, {\em Optical solitons: From Fibers to Photonic Crystals}, (Academic Press, 2003).
\bibitem{lederer} F. Lederer, G. I. Stegeman, D. N. Christodoulides, G. Assanto, M. Segev, and Y. Silberberg, Phys. Rep. {\bf 463}, 1 (2008).
\bibitem{tranBBDS} M. C. Tran and Tr. X. Tran, Chaos {\bf 32}, 073113 (2022).
\bibitem{talbot} H. F. Talbot, Philos. Mag. {\bf 9}, 401 (1836).
\bibitem{iwanow} R. Iwanow, D. A. May-Arrioja, D. N. Christodoulides, G. I. Stegeman, Y. Min, and W. Sohler, Phys. Rev. Lett. {\bf 95}, 053902 (2005).

\bibitem{tranresonant1} Tr. X. Tran and F. Biancalana, Phys. Rev. Lett. {\bf 110}, 113903 (2013).

\bibitem{tranresonant3} Tr. X. Tran, D. C. Duong, and F. Biancalana, Phys. Rev. A {\bf 89}, 013826 (2014).
\bibitem{bloch} F. Bloch, Z. Phys. {\bf 52}, 555 (1928).
\bibitem{blochZener} C. Zener, Proc. R. Soc. London Ser. A {\bf 145}, 523 (1934).
\bibitem{pertsch} T. Pertsch, P. Dannberg, W. Elflein, A. Br\"{a}uer, and F. Lederer, Phys. Rev. Lett. {\bf 83}, 4752 (1999).
\bibitem{lenz} G. Lenz, I. Talanina, and C. M. de Sterke, Phys. Rev. Lett. {\bf 83}, 963 (1999).
\bibitem{ghulinyan} M. Ghulinyan, C. J. Oton, Z. Gaburro, L. Pavesi, C. Toninelli, and D. S. Wiersma, Phys. Rev. Lett. {\bf 94}, 127401 (2005).
\bibitem{trompeter} H. Trompeter, T. Pertsch, F. Lederer, D. Michaelis, U. Streppel, and A. Br\"{a}uer, Phys. Rev. Lett. {\bf 96}, 023901 (2006).
\bibitem{zitterbewegung} F. Dreisow, M. Heinrich, R. Keil, A. T\"{u}nnermann, S. Nolte, S. Longhi, and A. Szameit, Phys. Rev. Lett. {\bf 105}, 143902 (2010).
\bibitem{trandirac2} Tr. X. Tran and D.C. Duong, Chaos {\bf 28}, 013112 (2018).
\bibitem{klein} O. Klein, Z. Phys. {\bf 53}, 157 (1929).
\bibitem{kleinLonghi} S. Longhi, Phys. Rev. B {\bf 81}, 075102 (2010).
\bibitem{kleinDreisow} F. Dreisow, R. Keil, A. T\"{u}nnermann, S. Nolte, S. Longhi, A. Szameit, Europhys. Lett. {\bf 97}, 10008 (2012).
\bibitem{tranKT1} M. C. Tran, Q. Nguyen-The, C. C. Do, and Tr. X. Tran, Phys. Rev. A {\bf 105}, 023523 (2022).
\bibitem{tranKT2} M. C. Tran, C. C. Do, and Tr. X. Tran, Ann. Physics {\bf 450}, 169241 (2023).
\bibitem{sauter} F. Sauter, Z. Phys. {\bf 69}, 742 (1931).
\bibitem{tranSauter} M. C. Tran and Tr. X. Tran, Ann. Physics {\bf 463}, 169624 (2024).
\bibitem{longhiPP} S. Longhi, Phys. Rev. A {\bf 81}, 022118 (2010).
\bibitem{dreisowPP} F. Dreisow, S. Longhi, S. Nolte, A. T\"{u}nnermann, and A. Szameit, Phys. Rev. Lett. {\bf 109}, 110401 (2012).
\bibitem{tranPP1} Tr. X. Tran, H. M. Nguyen, and D. C. Duong, Phys. Rev. A {\bf 105}, 032201 (2022).
\bibitem{tranPP2} Tr. X. Tran, H. M. Nguyen, and D. C. Duong, Ann. Physics {\bf 459}, 169528 (2023).
\bibitem{jackiw} R. Jackiw and C. Rebbi, Phys. Rev. D {\bf 13}, 3398 (1976).
\bibitem{tranjr4} Tr. X. Tran, J. Opt. Soc. Am. B {\bf 36}, 2559 (2019).
\bibitem{rayleigh} L. Rayleigh, Philos. Mag. {\bf 11}, 196 (1881).
\bibitem{winthrop} J. T. Winthrop and C. R. Worthington, J. Opt. Soc. Am. {\bf 55}, 373 (1965).
\bibitem{szwaykowski} P. Szwaykowski and V. Arrizon, Appl. Opt. {\bf 32}, 1109 (1993).
\bibitem{berry} M. V. Berry and S. Klein, J. Mod. Opt. {\bf 43}, 2139 (1996).
\bibitem{clauser} J. F. Clauser and J. P. Dowling, Phys. Rev. A {\bf 53}, 4587 (1996).
\bibitem{mehuys} D. Mehuys, W. Streifer, R. G. Waarts, and D. F. Welch, Opt. Lett. {\bf 16}, 823 (1991).
\bibitem{azana} J. Azana, Opt. Lett. {\bf 30}, 227 (2005).
\bibitem{robinett} R. W. Robinett, Phys. Rep. {\bf 392}, 1 (2004).
\bibitem{deng} L. Deng, E. W. Hagley, J. Denschlag, J. E. Simsarian, M. Edwards, C. W. Clark, K. Helmerson, S. L. Rolston, and W. D. Phillips, Phys. Rev. Lett. {\bf 83}, 5407 (1999).
\bibitem{chapman} M. S. Chapman, C. R. Ekstrom, T. D. Hammond, J. Schmiedmayer, B. E. Tannian, S. Wehinger, and D. E. Pritchard, Phys. Rev. A {\bf 51}, R14 (1995).
\bibitem{eberly} J. H. Eberly, N. B. Narozhny, and J. J. Sanchezmondragon, Phys. Rev. Lett. {\bf 44}, 1323 (1980).
\bibitem{kaplan} A. E. Kaplan, I. Marzoli, W. E. Lamb, and W. P. Schleich, Phys. Rev. A {\bf 61}, 032101 (2000).
\bibitem{breid} B. M. Breid, D. Witthaut, and H. J. Korsch, New J. Phys. {\bf 8}, 110 (2006).
\bibitem{longhiZener} S. Longhi, EuroPhys. Lett. {\bf 76}, 416 (2006).
\bibitem{dreisowZener} F. Dreisow, A. Szameit, M. Heinrich, T. Pertsch, S. Nolte, A. T\"{u}nnermann, and S. Longhi, Phys. Rev. Lett. {\bf 102}, 076802 (2009).
\bibitem{wen} J. Wen, Y. Zhang, and M. Xiao, Adv. Opt. Phot. {\bf 5}, 83 (2013).
\bibitem{sukhorukov1} A. A. Sukhorukov and Y. S. Kivshar, Opt. Lett. {\bf 27}, 2112 (2002).




\end{thebibliography}
\end{document}